\documentclass[fleqn,usenatbib]{mnras}

\usepackage[T1]{fontenc}

\DeclareRobustCommand{\VAN}[3]{#2}
\let\VANthebibliography\thebibliography
\def\thebibliography{\DeclareRobustCommand{\VAN}[3]{##3}\VANthebibliography}

\usepackage{graphicx}	% Including figure files
\usepackage{amsmath}	% Advanced maths commands
\usepackage{amssymb}	% Extra maths symbols
\usepackage{xcolor}
\usepackage[normalem]{ulem}
\usepackage{float}

\definecolor{Joao}{RGB}{102,0,0}
\definecolor{JoaoII}{RGB}{102,0,102}

\newcommand\rstrike{\bgroup\markoverwith{\textcolor{red}{\rule[0.5ex]{2pt}{0.4pt}}}\ULon}

\newcommand\rstrikeII{\bgroup\markoverwith{\textcolor{red}{\rule[0.5ex]{2pt}{0.4pt}}}\ULon}

\title[ADM constraints from subgiant asteroseismology]{On asymmetric dark matter constraints from the asteroseismology of a subgiant star}

\author[Rato et al.]{
João Rato$^{1}$\thanks{E-mail: jpedrorato@tecnico.ulisboa.pt}, José Lopes$^{1}$ and Ilídio Lopes$^{1}$
\\
$^{1}$Centro de Astrofísica e Gravitação - CENTRA, Departamento de Física, Instituto Superior T\'ecnico - IST, Universidade de Lisboa - UL, Av. Rovisco Pais 1,\\ P-1049-001 Lisboa, Portugal\\
}

\date{\today}

\pubyear{2021}

\usepackage{newtxtext,newtxmath}

\begin{document}
\label{firstpage}
\pagerange{\pageref{firstpage}--\pageref{lastpage}}
\maketitle

\begin{abstract}
The asteroseismic modelling of solar-like stars has proved to be valuable in constraining dark matter.
In this work we study for the first time the influence of asymmetric dark matter (ADM) in the evolution of a subgiant star (KIC 8228742) by direct comparison with observational data.
Both spectroscopic and seismic data are analysed with a new approach to the stellar calibration method, in which DM properties can also be considered as free inputs.
In another phase of this study, a calibrated standard stellar model (without DM) is used as the benchmark for DM models.
We find that the latter models consistently outperform the former for $10^{-40} \leq \sigma_\mathrm{SD} < 10^{-38}$ cm$^2$, hinting that the presence of ADM in stars of this type does not go against observations.
Moreover, we show that stellar seismology allows us to suggest exclusion limits that complement the constraints set by direct detection experiments.
Different seismic observables are proposed to study DM properties and $\Delta\Pi_\ell$ is found to be the most reliable, having the potential to build future DM exclusion diagrams.
This new methodology can be a powerful tool in the analysis of the data coming from the next generation of asteroseismic missions.
\end{abstract}


%%%%%%%%%%%%%%%%%%%%%%%%%%%%%%%%%%%%%%%%%%%%%%%%%%

%%%%%%%%%%%%%%%%% BODY OF PAPER %%%%%%%%%%%%%%%%%%

\section{Introduction} \label{sec:Introduction}

Since the dark matter (DM) hypothesis arose from the discrepancy between theory and the observation of the galaxies' rotation curves (Rubin et al. \citeyear{Rubin_1976}), direct detection experiments have been conducted in the hope of finding these elusive particles (e.g. Bertone \citeyear{Bertone_2010}; Undagoitia \& Rauch \citeyear{Undagoitia_2015} and references therein).
Although some constraints on the mass and cross section of interactions of DM particles with baryons have been set (e.g. Zyla et al. \citeyear{Zyla_2020}), no detection has yet been confirmed (e.g. Schumann \citeyear{Schumann_2019}).

Weakly Interacting Massive Particles (WIMPs) stand out as one of the primary candidates for DM (Bertone, Hooper \& Silk \citeyear{Bertone_2005}; Bertone \& Hooper \citeyear{Bertone_2018}). 
These particles have a non-negligible scattering cross section with baryons which is usually treated in two separate components: spin-dependent, $\sigma_{\mathrm{SD}}$, and spin-independent interactions, $\sigma_{\mathrm{SI}}$ (Barger, Keung \& Shaughnessy \citeyear{Barger_2008}).
For WIMP masses around $m_{\chi}\simeq5~\mathrm{GeV}$, recent upper limiting constraints on $\sigma_\mathrm{SD}$ (WIMP-proton interactions) have been placed at slightly below 5$\times10^{-38}$ cm$^2$ by PICASSO (Behnke et al. \citeyear{PICASSO_2017}) and at around 4$\times10^{-40}$ cm$^2$ by PICO-60 (Amole et al. \citeyear{PICO_2019}).
The XENON-100 experiment (Aprile et al. \citeyear{XENON_2016}) placed a limit for $\sigma_\mathrm{SD}$ at $\sim$10$^{-36}$ cm$^2$ for $m_\chi \sim 9$ GeV which was improved on by the XENON1T experiment (Aprile et al. \citeyear{XENON1T_2019}), reporting a limit at $\sigma_\mathrm{SD}\sim 4.7\times10^{-39}$ cm$^2$ for the same value of mass.
For $\sigma_\mathrm{SI}$ (WIMP-nucleon interactions), the upper limits for $m_\chi \sim 10$ GeV were found to be at around 5$\times$10$^{-41}$ cm$^2$ from PICASSO, at 2$\times$10$^{-43}$ cm$^2$ from PICO-60, and at 5$\times$10$^{-45}$ cm$^2$ from XENON1T (Aprile et al. \citeyear{Aprile_2019}).

In this study, we consider WIMPs in an asymmetric dark matter (ADM) scenario, in which the DM annihilation cross section is negligible.
Much like baryons in baryogenesis, DM asymmetry is hypothesised to have been produced in a process often called darkogenesis (e.g. Shelton \& Zurek \citeyear{Shelton_2010}).
Thus, in the ADM framework, the DM and anti-DM densities are unbalanced and make the present-day DM self-annihilation negligible.
This choice of framework is mainly interesting in the standpoint of the DM influence on stars: since DM self-annihilation does not occur, the number of DM particles inside a star will naturally be larger than it would be otherwise, making the star more sensible to its effects, which allows for a better study of DM phenomena.

Besides the already mentioned direct detection achievements, stars have also been used in the search for DM.
These endeavours have ranged from the study of solar models affected by DM (e.g. Lopes \& Silk \citeyear{Lopes_2010}; Taoso et al. \citeyear{Taoso}) to asteroseismic analysis (e.g. Casanellas \& Lopes \citeyear{Casanellas2013}; Lopes, Lopes \& Silk \citeyear{Lopes_2019b}) also including neutrino flux constraints (e.g. Turck-Chièze et al. \citeyear{Turck-Chieze_2012}).
Using stars and stellar models as an object of study of DM also has its shortcomings, which are mostly inherited from standard stellar modelling.
A notable example among these is the so-called solar composition problem.
Standard solar models using the most recent photospheric abundances (Asplund et al. \citeyear{Asplund_2009}, hereafter AGSS09) as inputs present a contradictory prediction of the Sun's internal structure when compared to high-precision results from helioseismology (e.g. Christensen-Dalsgaard \citeyear{Christensen_Dalsgaard_2002}; Bahcall, Serenelli \& Basu \citeyear{Bahcall_2005}).
This discrepancy between predictions coming from spectroscopy and helioseismology renders the determination of stellar properties (such as the sound speed profile) through stellar modelling problematic and affects not only the modelling of the Sun but also other stars since they rely on solar inputs for some quantities, like the relative metallicity $Z/X$.
In a recent discussion of this problem, Capelo \& Lopes (\citeyear{Capelo_2020}) have shown that measuring neutrino fluxes from the CNO cycle with a precision that could be achieved by the next generation of experiments could help resolve this issue.
While this problem can hinder the ability of using stellar modelling to probe DM properties, DM itself can also be an answer to the abundance problem since it introduces different physics in the interior of stars.
This idea was first suggested by Frandsen \& Sarkar (\citeyear{Frandsen_2010}) who reported that the presence of ADM in the Sun could resolve this issue.
Later in the same year, Taoso et al. (\citeyear{Taoso}) investigated this claim through numerical simulations, although not reaching the same conclusions.
Cumberbatch et al. (\citeyear{Cumberbatch_2010}) added to this discussion by noting that the effect of WIMP presence is not large enough to account for the discrepancies between observations and helioseismology.
However, they pointed out that this effect may still play a role and should be included in models.
Particularly, Lopes, Panci \& Silk (\citeyear{Lopes_2014}) proposed that the accretion of DM in the Sun's core could lead to a better agreement between helioseismic and neutrino data.
In a follow up with a more detailed analysis, Vincent, Scott \& Serenelli (\citeyear{Vincent_2015}) show that the solar abundance problem could be solved by the presence of a light ADM particle.

Asteroseismology has particularly been thoroughly exploited in an attempt to find constraints for the properties of DM while using stars as laboratories (Casanellas \& Lopes \citeyear{Casanellas_2010}).
The premise is fairly straightforward: by analysing oscillation frequencies of stars we can extract valuable information of their interior structure.
This is related to DM because the capture and subsequent accumulation of these particles via gravitational effects introduce an additional energy transport mechanism.
This new phenomenology can naturally lead to changes in the structure of the star, which can be probed via asteroseismic diagnostics (e.g. Casanellas \& Lopes \citeyear{Casanellas_2010}; Martins, Lopes \& Casanellas \citeyear{Martins_2017}).
Missions like CoRoT (Fridlund et al. \citeyear{CoRoT_2006}; Micher et al. \citeyear{CoRoT_2008}), \textit{Kepler} (Borucki et al. \citeyear{Kepler_2010a}; Koch et al. \citeyear{Kepler_2010b}) and TESS (Ricker et al. \citeyear{TESS}) have made progress in obtaining the oscillation frequencies of many main-sequence (MS), subgiant (SG) and red-giant (RG) stars with great precision, making it possible to study the asteroseismology of stars other than the Sun.
To take full advantage of this diversity of data, seismic diagnostics can be formulated for several stars in different stages of evolution, broadening the spectre of potential DM laboratories.
With the primary goal of discovering habitable extra-solar planets, the PLAnetary Transits and Oscillations of stars (PLATO) mission (Rauer et al. \citeyear{PLATO}), to be launched in 2026, will extend this effort and enable more precise studies by the determination of accurate stellar masses, radii, and ages from asteroseismic data.
The oscillation frequency measurements are expected to improve in precision upon those of \textit{Kepler} while also extending the catalogue to include brighter stars.
Thus, this will enable the study of the effects of DM on the stellar structure with both greater precision and for a considerably larger number of stars.

In the following section we focus on the asteroseismology of SG stars and important quantities for describing stellar oscillations are introduced.
We then describe the calibration methodology and the diagnostics used to infer on the quality of the calibrated models, in Section~\ref{sec:methods}.
In Section~\ref{sec:ADM} we address the interactions between ADM and stars.
Results are then shown in two separate parts.
In Section~\ref{sec:stellarmodels} we calibrate a SG star with and without ADM presence, where the stellar inputs are treated as free parameters and are thus optimised to better fit the observational data.
After that, in Section~\ref{sec:discussion}, we take on the work of the previous section by choosing the best no-DM model as benchmark to build a set of DM models with different properties and fixed standard stellar inputs.
We then use the diagnostics defined in Section~\ref{sec:methods} to classify this set of models and enquire about constraints on the properties of ADM.
Finally, conclusions and closing remarks are presented in the last section.

\section{Asteroseismology of subgiant stars}
\label{sec:astero}

The study of the impact of DM on the Sun and other MS sun-like stars using asteroseismology has allowed to constrain properties of different types of particle DM (e.g. Frandsen \& Sarkar \citeyear{Frandsen_2010}; Lopes et al. \citeyear{Lopes_2019b}).
In this work we focus on stars in a different stage of evolution -- the SG phase.
This stage follows the MS, i.e., after hydrogen burning ceases in the centre of the star and moves to a shell right above the helium ashes that compose the inert stellar core.

In particular, the object of our study is the subgiant KIC 8228742, a F9IV-V spectral type star (Molenda-Zakowicz et al. \citeyear{Molenda-Zakowicz_2013}) with a previously modelled mass of 1.27 M$_{\sun}$ (Metcalfe et al. \citeyear{Metcalfe_2014}).
Throughout this work, the observational constraints were taken from Chaplin et al. (\citeyear{Chaplin_2013}) for the spectroscopic parameters and from Appourchaux et al. (\citeyear{Appourchaux_2012}) for the oscillation frequencies, as published in the Asteroseismic Modelling Portal (AMP; Metcalfe \citeyear{AMP}).

Despite the experimental advances made by the aforementioned missions, SG stars are more difficult to find than MS or giant stars since that stage has a relatively shorter lifetime.
Another interesting aspect when studying the asteroseismology of SG stars is the lack of detected non-radial acoustic modes.
Usually, SG and RGs' most visible oscillations are gravity-dominated mixed modes (Hekker \& Mazumdar \citeyear{Hekker_2013}; Gai et al. \citeyear{Gai_2017}): due to the rapid core contraction, the gravity (g-) and acoustic (p-) mode trapping cavities are closer to each other when the stars move off the MS, which results in the coupling of the acoustic and gravity modes.
These are called mixed modes, which have p-mode characteristics in the convective stellar envelope and g-mode characteristics in the dense radiative stellar core.
However, KIC 8228742 exhibits many simple modes (or pure acoustic p-modes) like the Sun (Appourchaux et al. \citeyear{Appourchaux_2012}).
This is due to gravity-dominated mixed modes having lower amplitudes than pressure-dominated ones (Dupret et al. \citeyear{Dupret_2009}; Grosjean et al. \citeyear{Grosjean_2014}), so observing them is not always easy.
As such, we first direct our focus to the acoustic simple modes.

Since acoustic waves of low spherical degree ($\ell$) propagate throughout the entire star, their frequencies encode information of the whole stellar structure, spanning from the star's core to its surface.
Thus, as stated before, obtaining them is pivotal to understand the underlying physics of the internal structure of a star.
Naturally, quantities stemming from these frequencies can be defined to better probe the stellar structure.
The large frequency separation is defined as (e.g. Tassoul \citeyear{Tassoul_1980}; Lopes \& Turck-Chièze \citeyear{Lopes_1994})
\begin{equation}\label{eq:DeltaNu}
    \Delta\nu_{n,\ell} = \nu_{n,\ell}-\nu_{n-1,\ell} \simeq\langle\Delta\nu_{n,l}\rangle = \left(2\int_0^R\frac{dr}{c(r)}\right)^{-1},
\end{equation}

where $\nu_{n,\ell}$ denotes the frequency of the mode with radial order $n$ and spherical degree $\ell$, $r$ is the radial coordinate, $R$ is the total radius of the star and $c(r)$ represents the sound speed profile inside the star.
Thus, $\Delta\nu_{n,\ell}$ is deeply related to the sound speed profile of a star and is useful as a global measure of that quantity (Floranes, Christensen-Dalsgaard \& Thompson \citeyear{Floranes_2005}).

Additionally, a small frequency separation can also be defined as (Tassoul \citeyear{Tassoul_1980}; Lopes \& Turck-Chièze \citeyear{Lopes_1994})
\begin{equation}
    \delta\nu_{n,\ell} = \nu_{n,\ell} - \nu_{n-1,\ell+2},
\end{equation}

which is particularly sensitive to the thermodynamic conditions of the stellar core.

As discussed before, SG and giant stars often exhibit gravity dominated modes.
From these, the period separation $\Delta\Pi_\ell$ is a useful quantity to extract and, in the asymptotic limit, it is given by (Tassoul \citeyear{Tassoul_1980})
\begin{equation}\label{eq:DeltaPi}
    \Delta\Pi_\ell = \frac{2\pi^2}{\sqrt{\ell(\ell+1)}}\left(\int_{r_1}^{r_2}N\frac{dr}{r}\right)^{-1} = \frac{\Pi_0}{\sqrt{\ell(\ell+1)}},
\end{equation}

where $N$ is the Brunt-Väisälä (or buoyancy) frequency and $r_1$ and $r_2$ correspond to the boundaries of the g-mode cavity which extends through the radiative region of the star.
Since, in SG stars, $r_1$ coincides with the interface between the inner convective zone (if there is one) and the radiative region, it follows that $\Delta\Pi_\ell$ directly relates to the size of the convective core.
%From its definition we see that, for SG stars, $\Delta\Pi_\ell$ directly relates to the size of the convective core.

\section{Calibration and diagnostic methods}
\label{sec:methods}

\subsection{Observables and calibration}

Stellar modelling has been extremely helpful in allowing us to better understand the physics at play inside stars. MESA (Modules for Experiments in Stellar Astrophysics; Paxton et al. \citeyear{Paxton2011}, \citeyear{Paxton2013}, \citeyear{Paxton2015}, \citeyear{Paxton2018}, \citeyear{Paxton2019}), an open-source 1-D stellar evolution code, is a powerful tool in this regard.
By combining various modules that aim to precisely describe different stellar phenomenology, MESA allows the user to model a wide variety of stars, given a set of stellar parameters as inputs.

In our work, we take advantage of the full capabilities of MESA, with special emphasis on the \textit{astero} module (Paxton et al. \citeyear{Paxton2013}), as it governs calibrations.
Using this module as a starting point, we carry out a stellar model calibration process which allows for both seismic and spectroscopic calibrations by taking as inputs \{$M$, $Y_i$, [Fe/H]$_i$, $\alpha$, $f_\mathrm{ov}$\} (stellar mass, initial helium abundance, initial metallicity, mixing-length parameter and overshooting parameter, respectively) and then producing an evolutionary model that is compared head-to-head with observations.
This comparison is accomplished by computing a $\chi^2_\mathrm{star}$ value that has weighted contributions from both spectroscopic ($\chi^2_\mathrm{spec}$) and seismic ($\chi^2_\mathrm{seis}$) observables.
In this work we use the default 2/3 weight on the seismic contribution and 1/3 on the spectroscopic counterpart, i.e., $\chi^2_\mathrm{star} = 1/3\chi^2_\mathrm{spec} + 2/3\chi^2_\mathrm{seis}$ (Metcalfe et al. \citeyear{Metcalfe_2012}; Paxton et al. \citeyear{Paxton2013}).
The diagnostics $\chi^2_\mathrm{spec}$ and $\chi^2_\mathrm{seis}$ are quadratic deviations from the spectroscopic and seismic observational data, respectively, with their uncertainties taken into account
\begin{equation}\label{eq:chi2star}
    \chi^2_\mathrm{spec/seis} = \frac{1}{N}\sum_{i=1}^N\left(\frac{X_i^\mathrm{mod} - X_i^\mathrm{obs}}{\sigma_{X_i}}\right)^2,
\end{equation}

\noindent where $N$ is the number of parameters, $X_i^\mathrm{mod}$ and $X_i^\mathrm{obs}$ are the stellar model and observed values of the $i^\mathrm{th}$ parameter, respectively, with $\sigma_{X_i}$ being the observational uncertainty.
The observational parameters used in $\chi^2_\mathrm{spec}$ are \{$L$, $T_\mathrm{eff}$, [Fe/H]\} = \{4.57 $\pm$ 1.45 L$_{\sun}$, 6042 $\pm$ 84 K, -0.14 $\pm$ 0.09\} (luminosity, effective temperature and metallicity), and in $\chi^2_\mathrm{seis}$ is \{$\Delta\nu$\} = \{62.1 $\pm$ 0.13 $\mu$Hz\}.

As for the calibration procedure, we use a method that is commonly used throughout the literature (e.g. Deheuvels et al. \citeyear{Deheuvels_2016}; Capelo \& Lopes \citeyear{Capelo_2020}) which relies on minimising $\chi^2_\mathrm{star}$.
This is achieved through an automatised optimisation process which uses a direct search method -- the downhill simplex algorithm (Nelder \& Mead \citeyear{Simplex}) -- to find the optimal set of inputs that produce the group of outputs \{$L$, $T_\mathrm{eff}$, [Fe/H], $\Delta\nu$\} that are closest to their observed counterparts.
To account for DM effects, we extend the standard calibration process to also include the relevant DM parameters as inputs, which extends the input set to \{$M$, $Y_i$, [Fe/H]$_i$, $\alpha$, $f_\mathrm{ov}$\, $m_\chi$, $\sigma_\mathrm{SD}$\} while maintaining the same outputs and comparison strategy.

\subsection{Seismic ratio diagnostics}
\label{subsec:seismicratios}

While ensuring that a model is consistent with observations in terms of spectroscopy is valuable in itself, in most cases these parameters do not fully mirror what is happening in the stellar interior.
These are the situations where thoroughly analysing the oscillation frequencies of a star becomes a powerful diagnostic tool.

To build upon the calibration process described in the last section, we resort to an additional seismic diagnostic of the stellar interior based on the observational frequencies $\nu_{n,\ell}$.
Since in stars other than the Sun it is difficult to observe modes with $\ell>2$ due to partial cancellation (e.g. Aerts, Christensen-Dalsgaard \& Kurtz \citeyear{Aerts_2010}), we decide to use the ratio of the small to large separations $r_{02}$,
\begin{equation}\label{eq:r02}
    r_{02}(n) = \frac{\delta\nu_{n,0}}{\Delta\nu_{n,1}}.
\end{equation}

Equation~(\ref{eq:r02}) aims to give better insights on the stellar core -- where the most significant DM influence is expected -- since the near-surface effects that highly affect individual frequency separations nearly cancel out by computing this ratio.
This means that $r_{02}(n)$ is independent of the structure of the outer layers of a star and thus works as a probe into the stellar interior (Roxburgh \& Vorontsov \citeyear{Roxburgh_2003}).
Conveniently, following the reasoning carried out in the calibration process, we define an additional quantity to assess the seismic quality of a given stellar model in terms of the $r_{02}$ ratio:
\begin{equation}\label{eq:chi2r02}
    \chi^2_{r_{02}} = \sum_{n=14}^{22}\left[\frac{r_{02}^\mathrm{obs}(n)-r_{02}^\mathrm{model}(n)}{\sigma_{r_{02}^\mathrm{obs}}}\right]^2,
\end{equation}

where $r_{02}^\mathrm{obs}(n)$ and $r_{02}^\mathrm{model}(n)$ represent the ratio defined in eq.~(\ref{eq:r02}) computed using both the observed and model frequencies, respectively, while $\sigma_{r_{02}^\mathrm{obs}}$ stands for the observed uncertainty which is computed through error propagation from the individual frequencies' uncertainties.
For the star considered here, we can observe 32 modes with $\ell\leqslant2$ and $n$ running from 14 to 22, amounting to 9 different instances of $r_{02}$ in the sum defined in eq.~(\ref{eq:chi2r02}).
The resulting value of $\chi^2_{r_{02}}$ along side the other mentioned diagnostics will provide the classification of the models.

It should be noted that the decision of computing and analysing $\chi^2_{r_{02}}$ separately from the calibration process is a compromise dictated by computational constraints:
while ideally the parameter $r_{02}$ could be included in the calibration process, it would be too time consuming to do so, given that it implies computing the specific mode frequencies, in opposition to the seismic parameters in $\chi^2_\mathrm{seis}$ which are computed with asymptotic approximations such as eq.~(\ref{eq:DeltaNu}). This two-step process allows us (within the existing limitations) to confirm that we are indeed choosing the best model, based the spectroscopic quantities considered in the calibration and confirmed by the actual frequency modes contained in $r_{02}$.

\begin{center}
\begin{table*}
\caption{KIC 8228742 models. The 4 bottom models were calibrated with fixed pairs of ($m_\chi$, $\sigma_\mathrm{SD}$) while DM Calib allowed the two parameters to vary. The first 2 columns are the DM parameters and the 3 following columns are some of the input parameters of the calibration.
Both $f_\mathrm{ov}$ and $[\mathrm{Fe}/\mathrm{H}]_i$ are omitted because they share similar values between all models ($f_\mathrm{ov} \simeq 1.62\times10^{-2}$ and $[\mathrm{Fe}/\mathrm{H}]_i \simeq -0.14$).
The following columns correspond (from left to right) to: age, luminosity, total radius and the logarithms of the central temperature and central density. The $\chi^2$ used in calibration and diagnostics are also displayed. Finally, the period spacing is shown for $\ell= 1$ and $\ell=2$.}
\label{table:benchmark}
\setlength{\tabcolsep}{5.3pt}
\begin{tabular}{l|cc|ccc|ccccc|cc|cc}
\hline\\
Model    & $m_\chi$ & $\sigma_\mathrm{SD}$ & $M$ & $Y_i$ & $\alpha$ & age & $L$ & $R$ & $\log \left(\frac{T_c}{1~\mathrm{K}}\right)$ & $\log\left(\frac{\rho_c}{1~\mathrm{gcm}^{-3}}\right)$ & $\chi^2_\mathrm{star}$ & $\chi^2_{r_{02}}$ & $\Delta\Pi_1$ & $\Delta\Pi_2$ \\
    & (GeV) & ($10^{-36}$ cm$^2$) & (M$_{\sun}$) & & & (Gyrs) & (L$_{\sun}$) & (R$_{\sun}$) & & & ($10^{-3}$) & & (s) & (s) \\\hline \vspace{0.3em}
SSG & - & - & 1.2565 & 0.244 & 1.403 & 4.52 & 4.254 & 1.884 & 7.42 & 2.53 & 5.735 & 27.8 & 1636 & 945 \\ \vspace{0.3em}
DM Calib & 9.12 & 2.32 & 1.2517 & 0.227 & 1.467 & 5.03 & 4.270 & 1.886 & 7.32 & 2.95 & 5.287 & 29.2 & 548 & 316 \\
DM A & 6.00 & 10$^{-3}$ & 1.2580 & 0.244 & 1.406 & 4.50 & 4.263 & 1.884 & 7.42 & 2.53 & 5.586 & 26.7 & 1642 & 948 \\
DM B & 6.00 & 10$^{-1}$ & 1.2591 & 0.243 & 1.407 & 4.50 & 4.264 & 1.885 & 7.42 & 2.53 & 5.486 & 37.8 & 1644 & 949 \\
DM C & 5.00$^a$ & 1 & 1.2516 
& 0.227 & 1.480 & 5.06 & 4.268 & 1.887 & 7.32 & 2.96 & 5.245 & 20.0 & 543 & 313 \\
DM D & 6.00 & 10 & 1.2590 & 0.229 & 1.403 & 4.76 & 4.339 & 1.884 & 7.32 & 2.96 & 17.879 & 64.3 & 564 & 326 \\
\hline
\multicolumn{15}{l}{\textit{Note.}\footnotesize$^a$ A slightly different mass value was used in this case due to model convergence limitations.}\\
\end{tabular}
\end{table*}
\end{center}

\section{Asymmetric dark matter interactions with a star}
\label{sec:ADM}

%In this work we explore the impact of ADM on KIC 8228742.
%For that, we need to treat the interactions of this type of matter with stars.
In the case of ADM, the interactions between DM particles and a given star are limited to capture, evaporation and energy transport, all due to scattering with the baryonic matter that constitutes the stellar plasma.
For the models considered in this work, it is safe to neglect evaporation -- the process in which DM particles that were already trapped inside the star scatter to velocities larger than the local escape velocity -- since it has been found that, for sun-like stars, the DM mass above which evaporation is negligible is close to 3.3 GeV (e.g. Gould \citeyear{Gould_1990}; Kouvaris \citeyear{Kouvaris_2015}), and here we explore larger mass values of ADM particles.
Additionally, as stated before, the annihilation cross section of ADM particles is negligible and, thus, the process that defines the number of DM particles inside the star is the capture.
This process, which consists in the gravitational trapping of DM particles from the galactic halo, is mainly defined by the DM mass, the DM--nucleon cross-section, the gravitational potential of the star and the local DM density. 
In our case, KIC 8228742 is just 0.17 kpc away from the Sun, therefore we assume a DM density corresponding to what is found in the solar neighbourhood, $\rho_\chi=0.38$ GeV/cm$^3$ (Catena \& Ullio \citeyear{Catena_2010}).

After accumulating and thermalising within the star, captured particles interact with baryons in the stellar interior.
These scatterings will create an additional mechanism for transporting energy, adding an extra term to the standard equation of energy transport in stars.
In most model-independent DM studies it is usual to assume the already mentioned $\sigma_\mathrm{SI}$ and $\sigma_\mathrm{SD}$ effective constant cross sections to describe the ADM particles' interactions with the baryons in the stellar plasma.
In this work, we focus on SD interactions, which, in the case of solar-like stars, correspond mostly to scatterings with hydrogen.
Previous studies regarding the impact of ADM in the Sun showed that due to the flux of energy carried outwards from the innermost regions of the star by ADM particles with a mass of 7 GeV and $\sigma_\mathrm{SD}\simeq 10^{-36}~\mathrm{cm}^2$, the stellar core exhibited a decrease in temperature when compared to the standard case, while the reverse happened with the (baryonic) density (Taoso et al. \citeyear{Taoso}; Iocco et al. \citeyear{Iocco_2012}, the latter with higher $\rho_\chi$).
This effect, whose intensity is naturally related to the interaction cross-section, is in fact counter intuitive given that any transport of energy away from the core should lead to its contraction, which would in turn lead to an increase in temperature.
This however is not the case, since the energy transported away by the ADM particles is of a higher order of magnitude than the one released by the core contraction, countering its heating effects.

Another well-known consequence of the extra energy transport by DM is the suppression of convection -- generally in the centre of the star -- which was firstly proposed by Renzini (\citeyear{Renzini}) and Bouquet \& Salati (\citeyear{Bouquet}) and later studied by Casanellas \& Lopes (\citeyear{Casanellas2013}) and Casanellas, Brandão \& Lebreton (\citeyear{Casanellas2015}), particularly for stars with masses between 1.1 and 1.3 M$_{\sun}$.
This suppression is directly related to the decrease in the temperature gradient, which prevents the arise of convection as it would in cases where there is no energy transport by DM.

In this work, to study the effects of the interactions between ADM and the stellar plasma during the evolution of star KIC 8228742 we modified the MESA stellar evolution code to include the processes described above, namely capture and energy transport.
We consider DM capture as described by Gould (\citeyear{Gould1987}), and energy transport is computed taking into account the numerical results by Gould \& Raffelt (\citeyear{gould1990a}).
During the evolution, the capture rate is computed at each time step, and total number of ADM particles inside the star is updated accordingly.
This information is then used to compute the extra energy term, which is fed to the usual set of differential equations that govern stellar evolution.

\section{Stellar Models}\label{sec:stellarmodels}

\subsection{Standard stellar model of the subgiant KIC 8228742}

Using the calibration methods described in Section~\ref{sec:methods} with all five inputs 
\{$M$, $Y_i$, [Fe/H]$_i$, $\alpha$, $f_\mathrm{ov}$\} as free parameters, we obtain several stellar models with no DM interactions.  % in an 8 core 3.6 GHz CPU.
From these models, the best one in terms of $\chi^2_\mathrm{star}$ was chosen as the benchmark model for future analysis and is henceforth also referred to as Standard Subgiant (SSG) model.
The resulting parameters are shown in the first row of Table \ref{table:benchmark}, where $\chi^2_{r_{02}}$ is also included.
It should be noted that directly comparing $\chi^2_\mathrm{star}$ with $\chi^2_{r_{02}}$ is misleading, as their definitions and normalisation are different.

%TABLE WAS HERE

In comparison with other models found in the literature, the SSG model's parameters fall well within the limits proposed in most works (e.g. Bellinger et al. \citeyear{Bellinger_2019}) with the exception of the initial metallicity and mixing-length parameters found in Verma et al. (\citeyear{starmodel2}).
This does not amount to a large discrepancy and thus the model is accepted to be a good reference model.
Additionally, the SSG model exhibits a convective core during the MS that extends up to 0.065 R$_{\sun}$ and showcases a helium core (surrounded by a hydrogen shell) at the end of the evolution (see Fig.~\ref{fig:abundance}), which is the expected structure for stars in this stage of evolution, with these values of stellar mass (e.g Hurley, Pols \& Tout \citeyear{Hurley_2000}; Salaris \& Cassisi \citeyear{Salaris_2005a}, \citeyear{Salaris_2005b}, \citeyear{Salaris_2005c}).

\subsection{Comparison of dark matter models with a standard stellar model}

A valuable asset of the improved calibration method considered in this work is that it allows for the DM properties to be treated as free parameters in the calibration.
In this sense, we allow the algorithm to vary the values of the ADM particles' mass in between 4 and 12 GeV and the spin-dependent cross section in between 10$^{-40}$ and 10$^{-35}$ cm$^2$ since this is both included in the region of the parameter space currently being probed by DM direct detection experiments and also the region that produces greater effects on stars (e.g. Casanellas \& Lopes \citeyear{Casanellas2013}; Martins et al. \citeyear{Martins_2017}).
In fact, during the development of this work, the results from the PICO-60 experiment (Amole et al. \citeyear{PICO_2019}) -- already presented in Section~\ref{sec:Introduction} -- were published and greatly challenged our choice of DM parameters (visible in the purple curves in Figs.~\ref{fig:chi2star2}-\ref{fig:chi2total}).
Nevertheless, we notice that since this region of the parameter space needs to be independently confirmed by other direct DM detection experiments, this work can contribute to the discussion.
In particular, by studying how DM affects the evolution of stars and by helping to constrain its properties in an independent way of experimental detectors.

As before, the standard stellar inputs shown in the previous section were treated as free parameters, so it is expected that the set of optimal parameters is different than the values shown in the first row of Table \ref{table:benchmark}.
Taking into account the DM phenomenology described in Section~\ref{sec:ADM}, we carried out the optimisation process, from which we retrieved the best model (i.e., with the lowest $\chi^2_\mathrm{star}$).
This model (DM Calib) found an optimal DM particle with a mass of $m_\chi = 9.12$ GeV and a spin-dependent interaction cross section of $\sigma_\mathrm{SD} = 2.32\times10^{-36}$ cm$^2$, which is within the limits of the XENON-100 experiment mentioned in Section~\ref{sec:Introduction}.
The fact that this model is calibrated, by definition, means that it is bound to be in agreement with the corresponding observations. However, it is interesting to note that the best agreement -- within the considered parameter range -- occurs for the aforementioned values of $m_\chi$ and $\sigma_\mathrm{SD}$, even though they fall inside the excluded area of other more recent experiments (e.g. Amole et al. \citeyear{PICO_2019}).

\begin{figure*}
    \centering\includegraphics[clip, trim=2.5cm 0cm 2cm 0cm, width=\linewidth]{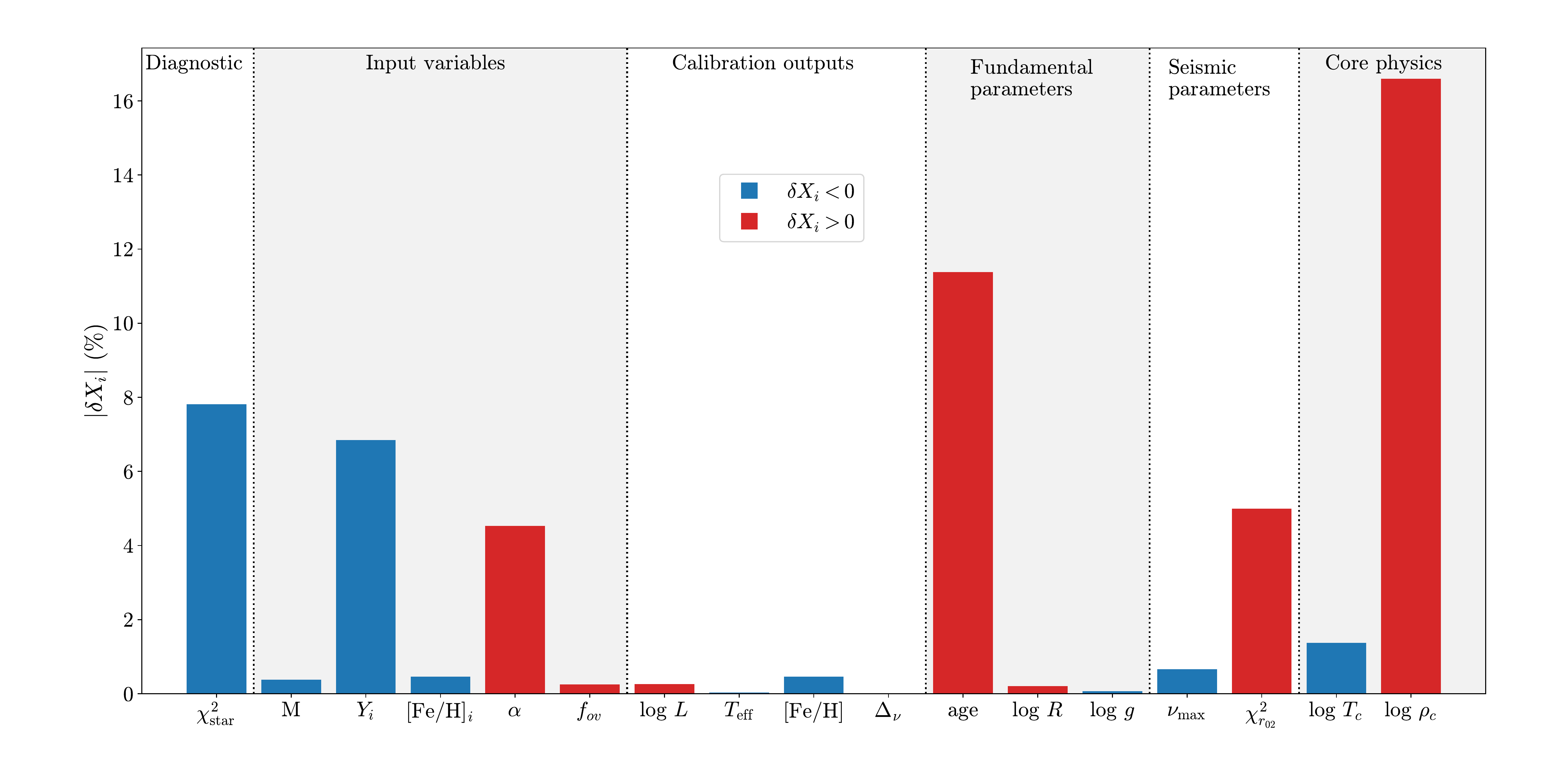}
    \vspace*{-10mm}\caption{Direct comparison of the DM Calib and Standard Subgiant models: the percentage variations were recorded relatively to the SSG model (see Table \ref{table:benchmark}). As some variations are negative, we use blue for negative values and red otherwise.}
    \label{fig:histogram_DM_noDM_comparison}
\end{figure*}

\begin{figure}
    \includegraphics[clip, trim=0.5cm 0cm 0cm 0.5cm, width=\linewidth]{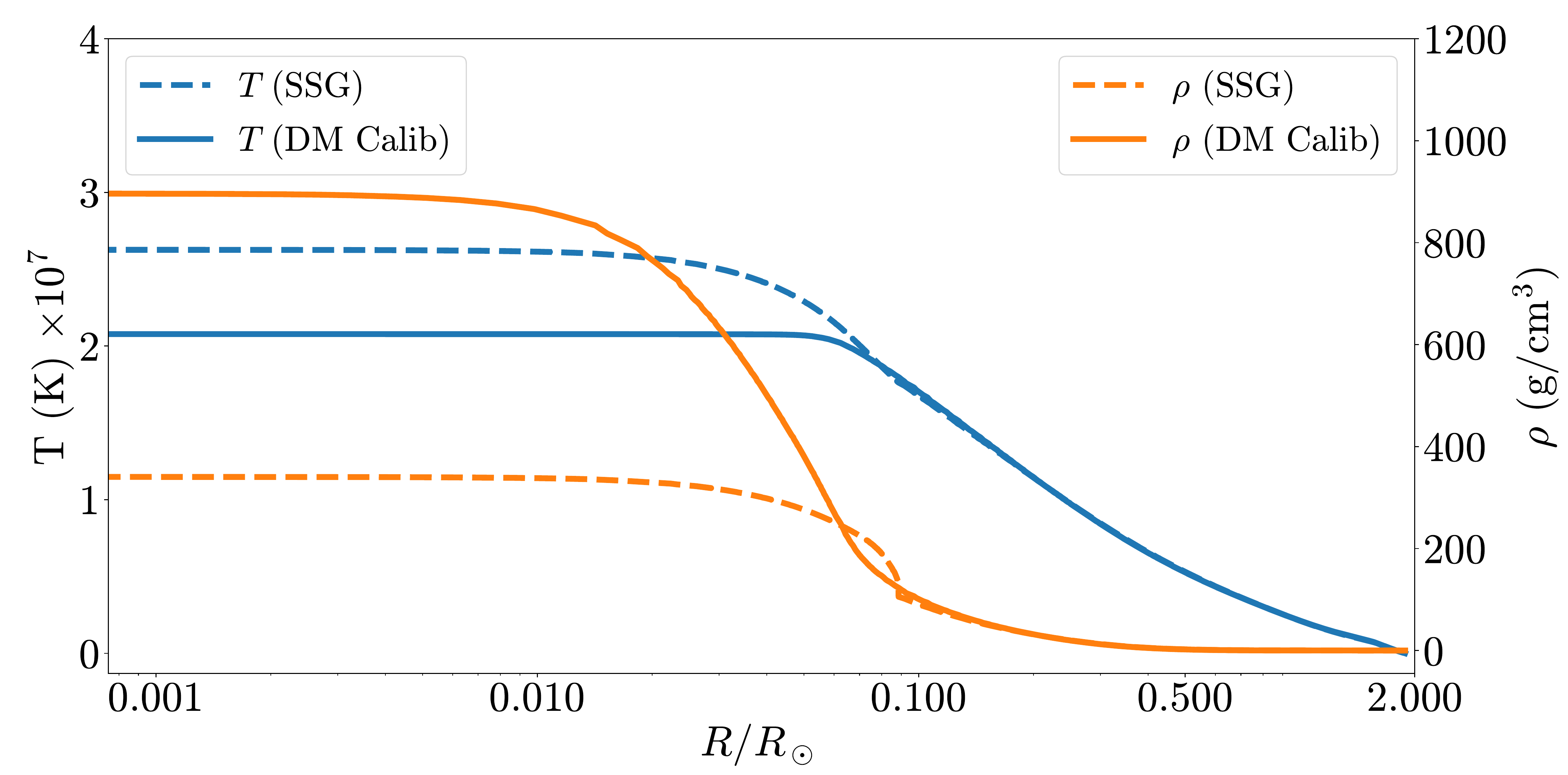}
    \vspace*{-5mm}\caption{Temperature (left axis) and baryonic density (right axis) profiles of the SSG and DM Calib stellar models (see Table \ref{table:benchmark}).}
    \label{fig:T_rho}
\end{figure}

\begin{figure}
    \includegraphics[clip, trim=1.5cm 0cm 1.5cm 0cm,width=\linewidth]{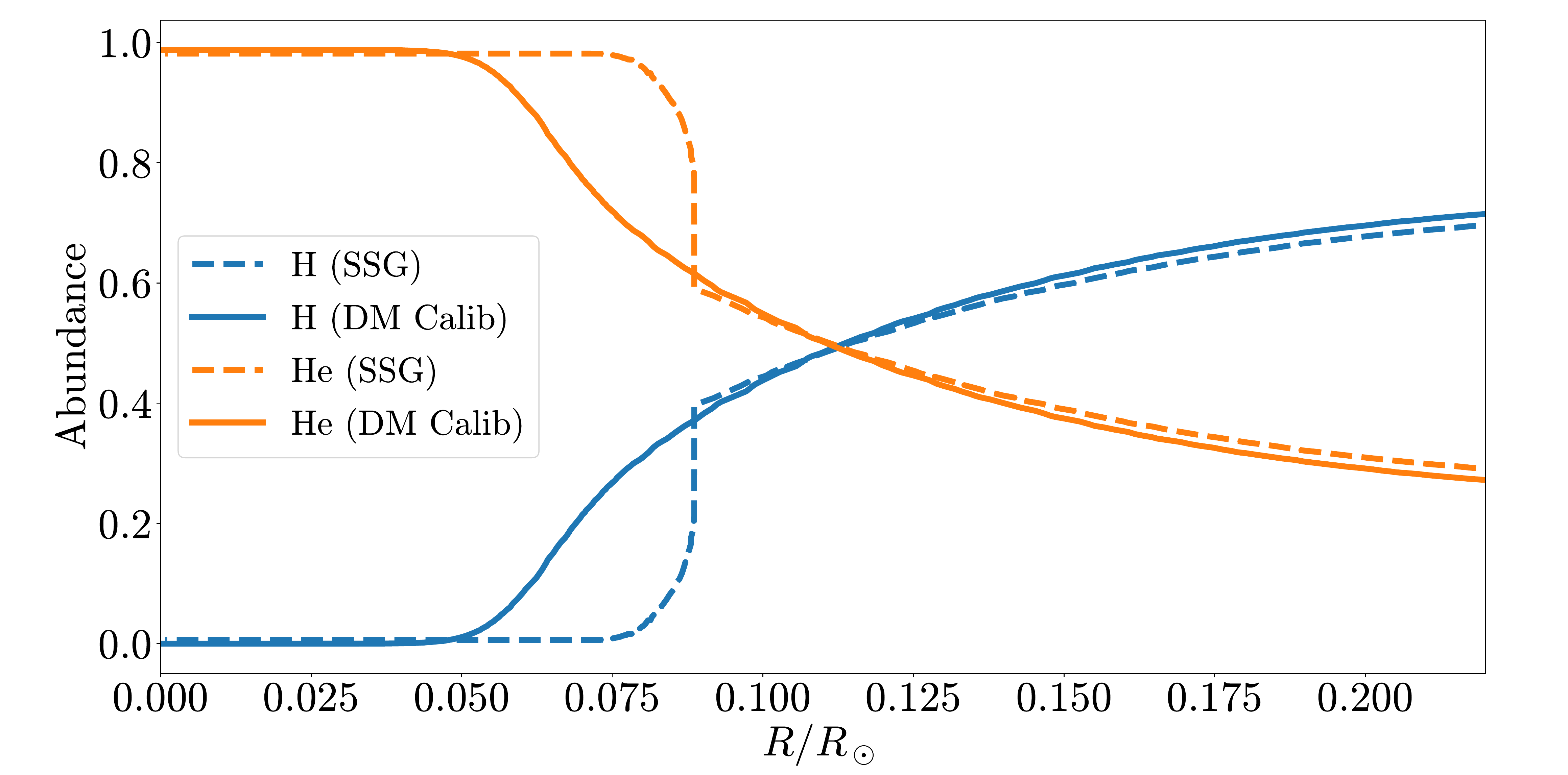}
    \vspace*{-5mm}\caption{Hydrogen and helium abundances of the SSG and DM Calib stellar models (see Table \ref{table:benchmark}).}
    \label{fig:abundance}
\end{figure}

\begin{figure}
    \includegraphics[clip, trim=0.5cm 0cm 1cm 1cm, width=\linewidth]{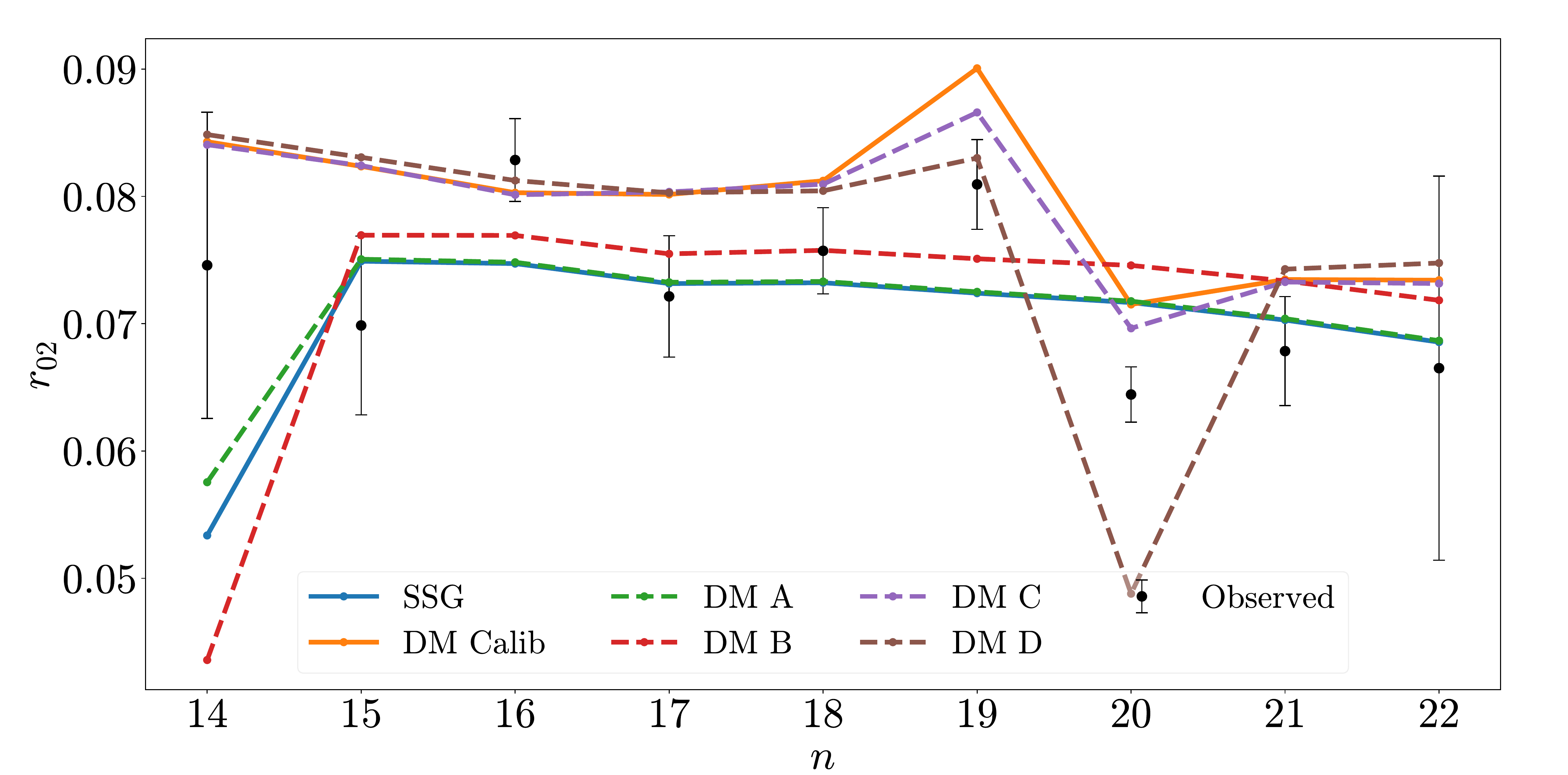}
    \vspace*{-5mm}\caption{$r_{02}$ ratios (see eq.~(\ref{eq:r02})) for the models in Table \ref{table:benchmark} compared to observations. The dashed lines represent the 4 models with fixed values of ($m_\chi$, $\sigma_\mathrm{SD}$).}
    \label{fig:r02}
\end{figure}

As expected, Fig.~\ref{fig:histogram_DM_noDM_comparison} shows that the optimal inputs changed with respect to the parameters obtained in the SSG model, some more drastically than others, as is the case of the initial Helium abundance $Y_i$ and the mixing-length parameter $\alpha$.
In terms of model outputs, the inclusion of the DM effects and parameters in the calibration process made the stellar age change in around 12\% while the most noticeable difference is the 16\% increase in the logarithm of the central density, which translates to a factor of around 2.5 in the central density itself.
This discrepancy is best viewed by comparing the respective profiles shown in Fig.~\ref{fig:T_rho}, where it is visible that the star forms an isothermal core at a lower temperature, which is consistent with other results for the Sun and other sun-like stars (e.g. Lopes \& Silk \citeyear{Lopes_2002}; Taoso et al. \citeyear{Taoso}; Iocco et al. \citeyear{Iocco_2012}).
Such is also the case of the baryonic density profile, which displays an increase in the innermost regions of the star when considering ADM capture and interactions.
One important aspect to remember is the fact that, by only considering spin-dependent couplings, the DM-stellar matter scatterings are practically reduced to DM-hydrogen interactions.
This means that, in fact, we neglect most of the DM energy transport that occurs during the actual SG phase, given that in these stars the region inhabited by ADM particles, i.e., the core, is mostly devoid of hydrogen.
Thus, the DM signatures shown here – and the resulting departures from the standard non-DM models – are in fact a consequence of DM interactions that occurred mainly during the MS.
Therefore, the comparison with different studies in the literature about similar effects in MS stars is reasonable since we are looking at the remnants of ADM interactions during the MS phase through a SG star.
It should also be noted that although we are not comparing the SSG and DM Calib models at the same age (nor do they have the same standard stellar inputs), as is the usual practice, the comparison is still of interest (and thus this effect is still expected) since it is made between two calibrated models of the same star, meaning that they are spectroscopically similar. %(i.e., they have approximately the same luminosity and effective temperature)
Nevertheless, one could argue that the change in the star's age could be the driving factor of the differences found between the two models.
But, in fact, that is not the case:
by analysing the same profiles of the two models for the same age (for example, at t = 4.52 Gyrs, see Table \ref{table:benchmark}) we confirm that the two distinct regimes are still present and identical to what is seen in Figs.~\ref{fig:T_rho} \& \ref{fig:abundance}.

Unlike the Standard model of this star, DM Calib did not exhibit a convective core.
In fact, the suppression of the convective core is a recurring feature of DM influence on stars as mentioned in Section~\ref{sec:ADM}.
Furthermore, by studying the star chemical profiles (Fig.~\ref{fig:abundance}) we see that while hydrogen is completely exhausted in the inner regions of the stellar core, there is a smooth increase in the hydrogen abundance (vice-versa for helium), instead of the sharp variation which is usual in stars with $M\simeq1.3~M_{\sun}$.
This is a direct consequence of the stellar core being radiative, as opposed to convective, during the main-sequence: the arise of core convection during the MS promotes the homogenisation of the chemical species in the central regions of the star, and thus the exhaustion of hydrogen that characterises the end of the MS occurs everywhere within the convective zone -- instead of locally in centre of the star.
%---leaving behind a discontinuity in the chemical abundances at the erstwhile convective boundary.

One other effect that is found in this model is the extension of the MS lifetime (Lopes \& Lopes \citeyear{Lopes_2019}; Raen et al. \citeyear{Raen_2021}).
While the star with DM has a radiative core during the largest part of its MS lifetime, and thus nuclear burning is limited to the local hydrogen supply, the decrease in central temperature slows down the hydrogen burning rate thus extending the MS lifetime.
The age of the model itself may be another indicator of this effect since the best agreements with observations (which are the goals of a successful calibration) were found to be at a later stellar age than in the standard case.

In addition to the model DM Calib, we decided to repeat the calibration process accounting for DM effects, but instead with pre-defined fixed values of $m_\chi$ and $\sigma_\mathrm{SD}$.
Stellar DM models from A to D are the best models for each of the ($m_\chi$, $\sigma_\mathrm{SD}$) pairs showcased in Table \ref{table:benchmark}.
The first conclusion to be taken is that DM Calib, which produced an optimal pair of ($m_\chi$, $\sigma_\mathrm{SD}$), does not have the lowest $\chi^2_\mathrm{star}$.
This means that the full calibration optimisation method most likely hit a local minimum around the displayed values.

A closer inspection at the remaining columns of Table \ref{table:benchmark} reveals that the SSG model, DM A and DM B all share similar outputs.
The same happens for models DM Calib, C and D, where the latter slightly deviates from the rest, leading to a $\chi^2_\mathrm{star}$ of an higher order of magnitude.
These two distinct regimes are expected once we take into account the mass and cross-section values of the DM impacted models: for $\sigma_\mathrm{SD} \geqslant 10^{-36}$ cm$^2$ the effects from ADM interactions have a noticeable impact on the star, while for smaller values of $\sigma_\mathrm{SD}$ the effects are mostly negligible.
We also confirm the prevalence of the two regimes when drawing profiles similar to Figs.~\ref{fig:T_rho} and \ref{fig:abundance} for the remaining models, i.e., the curves of Standard, DM A and DM B have similar behaviour between themselves whilst the remaining models follow the signature of DM Calib.

~\\

\vspace*{-7mm}
Finally, it is interesting to note that some DM models have a lower $\chi^2_\mathrm{star}$ than the SSG model.
For DM Calib, the decrease in almost 8\% of this quantity is substantial and the fact that DM A, B and C also represent an improvement on the Standard model's value reinforces the argument that the existence of DM is not incompatible with the current observational data for this star.
However, the $\chi^2_{r_{02}}$ diagnostic increased in about 4\% for DM Calib (even more for DM B) which hints towards the fact that the model might have fallen victim to an equivalent of overfitting.
This means that, since the optimisation is done with respect to $\chi^2_\mathrm{star}$, other parameters of the star might have been affected to achieve a better performance in that specific diagnostic.
Either way, the deviation on $\chi^2_{r_{02}}$ is not as significant as in the previous diagnostic.
To better infer on the $\chi^2_{r_{02}}$ discrepancy, we compute $r_{02}(n)$ as shown in Fig.~\ref{fig:r02}.

The results for $r_{02}(n)$ show that although the DM Calib ratio values deviate more from the observational average than the SSG's, the behaviour pattern is similar to that of the observed $r_{02}$ (particularly around $n \sim 19-20$).
Fig.~\ref{fig:r02} also shows that model DM C has the same regime as DM Calib, but slightly closer to the observation values, which is reflected in its smaller $\chi^2_{r_{02}}$.
Again, the two regimes are distinctively visible, with DM B being somewhat of an intermediate model.
This is also expected since in this model $\sigma_\mathrm{SD}$ is in-between that of DM A and DM C, which showcase each one of the two different regimes.

\section{Asteroseismic analysis} \label{sec:discussion}

\subsection{Probing the parameter space \texorpdfstring{$m_\chi-\sigma_\mathrm{SD}$}{ms}}

\begin{figure}
    \includegraphics[clip, trim=2.5cm 0cm 1.5cm 0cm, width=\linewidth]{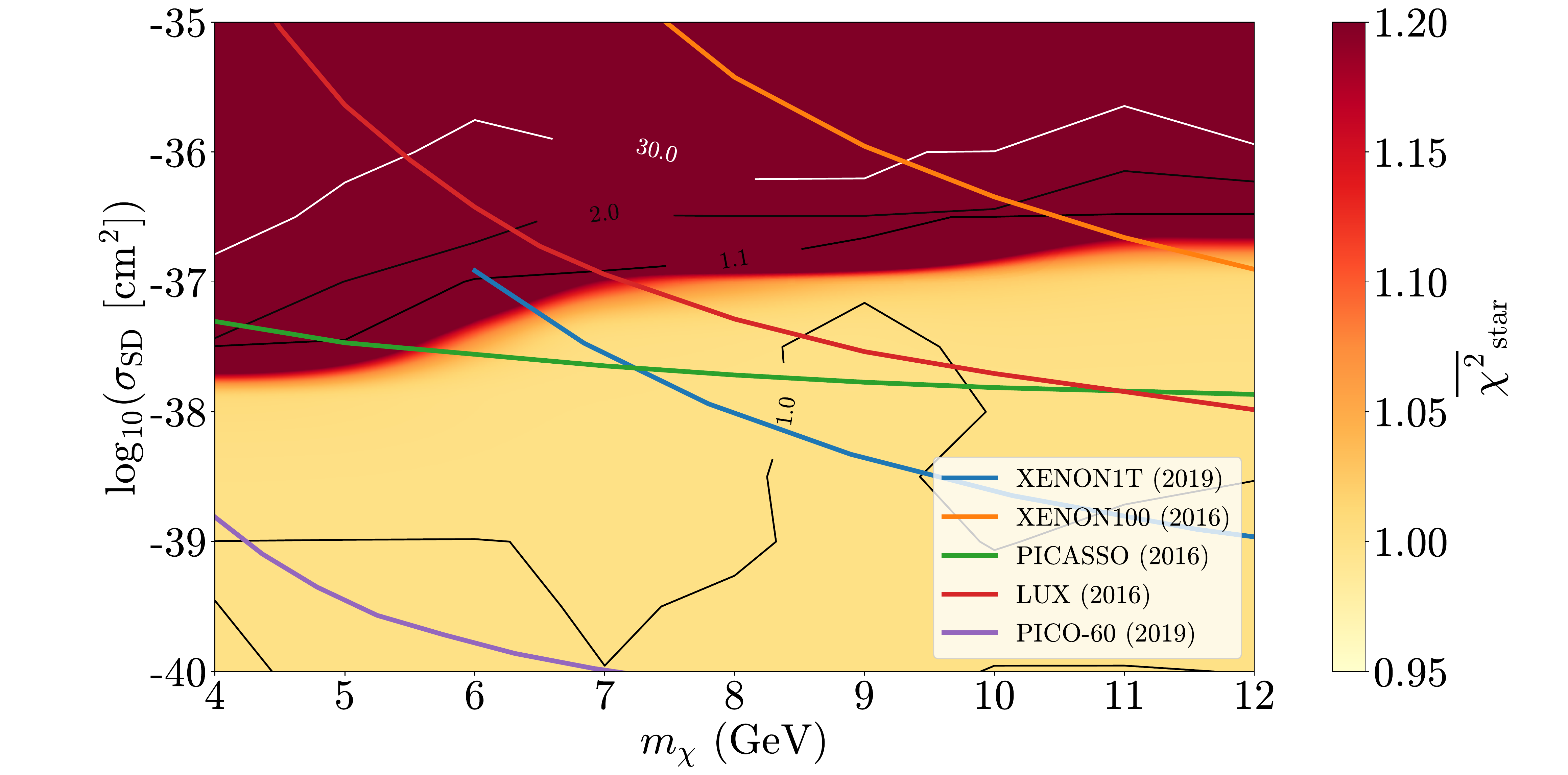}
    \caption{Contours for $\overline{\chi}^2_\mathrm{star}$ (see eqs.~(\ref{eq:chi2star}) and (\ref{eq:chi2starnorm})): lighter shades represent lower $\chi^2$ and thus better models for this specific diagnostic. The contour for $\overline{\chi}^2_\mathrm{star}=1$ represents models as good as the SSG model. Direct detection experiment upper bound limits from XENON1T (Aprile et al. \citeyear{XENON1T_2019}), XENON-100 (Aprile et al. \citeyear{XENON_2016}), LUX (Akerib et al. \citeyear{LUX}), PICASSO (Behnke et al. \citeyear{PICASSO_2017}) and PICO-60 (Amole et al. \citeyear{PICO_2019}) are shown as solid lines.}
    \label{fig:chi2star2}
\end{figure}

Using the input stellar parameters of the SSG model as benchmark (see Table \ref{table:benchmark}) and enabling DM interactions, we decide to explore the sensibility of the models in DM parameter space.
This is achieved by computing 100 models in a $m_\chi-\sigma_\mathrm{SD}$ grid with fixed DM parameters within the range 4 $\leqslant$ $m_\chi$ $\leqslant$ 12 GeV and 10$^{-40}$ $\leqslant$ $\sigma_\mathrm{SD}$ $\leqslant$ $10^{-35}$ cm$^2$.
As before, this range of values was chosen in agreement with the DM parameter space usually explored in the literature (e.g. Martins et al. \citeyear{Martins_2017}). Each input group in this grid was then used to create a model (i.e., one single model, differently from the optimisation process discussed in the previous sections) for which $\chi^2_\mathrm{star}$ was computed, allowing for the drawing of contour plots showcasing the parameter region of interest and corresponding DM parameters (Fig.~\ref{fig:chi2star2}).
A normalised $\overline{\chi}^2_\mathrm{star}$ was defined as:
\begin{equation}\label{eq:chi2starnorm}
    \overline{\chi}^2_\mathrm{star} = \frac{\chi^2_{\mathrm{DM}}}{\chi^2_{\mathrm{SSG}}},
\end{equation}
where $\chi^2_{\mathrm{DM}}$ corresponds to the $\chi^2_\mathrm{star}$ of each model taking DM into account and, likewise, $\chi^2_{\mathrm{SSG}}$ is that same value for the Standard model (in this case $\chi^2_{\mathrm{SSG}}$ = 5.735$\times10^{-3}$, see Table \ref{table:benchmark}).

On the one hand, it is visible that the overall tendency is for stellar models with lower interaction cross-section to agree better with observations.
On the other hand, there are some models with higher $\sigma_\mathrm{SD}$ that do not converge, meaning that the benchmark input parameters coupled with the given DM quantities cannot converge to an acceptable solution of the stellar evolution differential equations.
This happened for models with $\sigma_\mathrm{SD}$ larger than $4\times10^{-36}$ cm$^2$, where although not explicitly shown in the figure, the $\chi^2$ values rose to orders of magnitude of 10$^2$.
This relation between the cross-section and $\overline{\chi}^2_\mathrm{star}$ is somewhat expected since the lower the $\sigma_\mathrm{SD}$ the smaller the influence of DM is on the stellar structure.
Thus, the lower region of the grid performs better than the upper region since it naturally tends to the SSG case.
However, it is still worth noting that for $\sigma_\mathrm{SD}$ as high as 10$^{-37}$ cm$^2$ some models seem to perform well.

Furthermore, by looking at the contour line that defines models as good as the SSG model (at $\overline{\chi}^2_\mathrm{star} = 1$, see Fig.~\ref{fig:chi2star2}), it is visible that a large portion of the 100 DM stellar models outperform it.
This means that most DM models with the same inputs as the SSG but with $m_\chi$ between 4 and 12 GeV and $\sigma_\mathrm{SD}$ between 10$^{-40}$ and $4\times10^{-39}$ cm$^{2}$ fit the spectroscopy and large frequency separation observations better than the best performing no-DM model.
It should be noted that despite the fact that $\sigma_\mathrm{SD} = 10^{-40}~\mathrm{cm}^2$ is a hard limit, i.e., it was chosen by default, that does not mean that the improvement in $\overline{\chi}^2_\mathrm{star}$ is observed for any $\sigma_\mathrm{SD}$ below this value.
In fact, we also obtained models below the minimum cross-section considered in Fig.~\ref{fig:chi2star2} (e.g. $\sigma_\mathrm{SD} = $ 10$^{-42}$ cm$^2$) and observed that the $\chi^2_\mathrm{star}$ diagnostic again tended to the SSG value (i.e. $\overline{\chi}^2_\mathrm{star} = 1$), which is expected since DM is less influential for lower interaction cross-sections.
Additionally, we found a model near $m_\chi = 9$ GeV and $\sigma_\mathrm{SD} = 3\times10^{-38}$ cm$^2$ that exhibits the lowest $\chi^2_\mathrm{star}$ of the set.

It is interesting, however, to study the performance of the best performing model and all the others under the $\chi^2_{r_{02}}$ diagnostic defined in Section~\ref{subsec:seismicratios}.
To achieve that, the $r_{02}$ ratio was computed for all models in the grid and compared to observations, i.e., the ratio that was computed with the 32 frequencies observed in the star.
After that, $\chi^2_{r_{02}}$ was computed and yet again plotted in contours.

\begin{figure}
    \includegraphics[clip, trim=2.5cm 0cm 1.5cm 0cm, width=\linewidth]{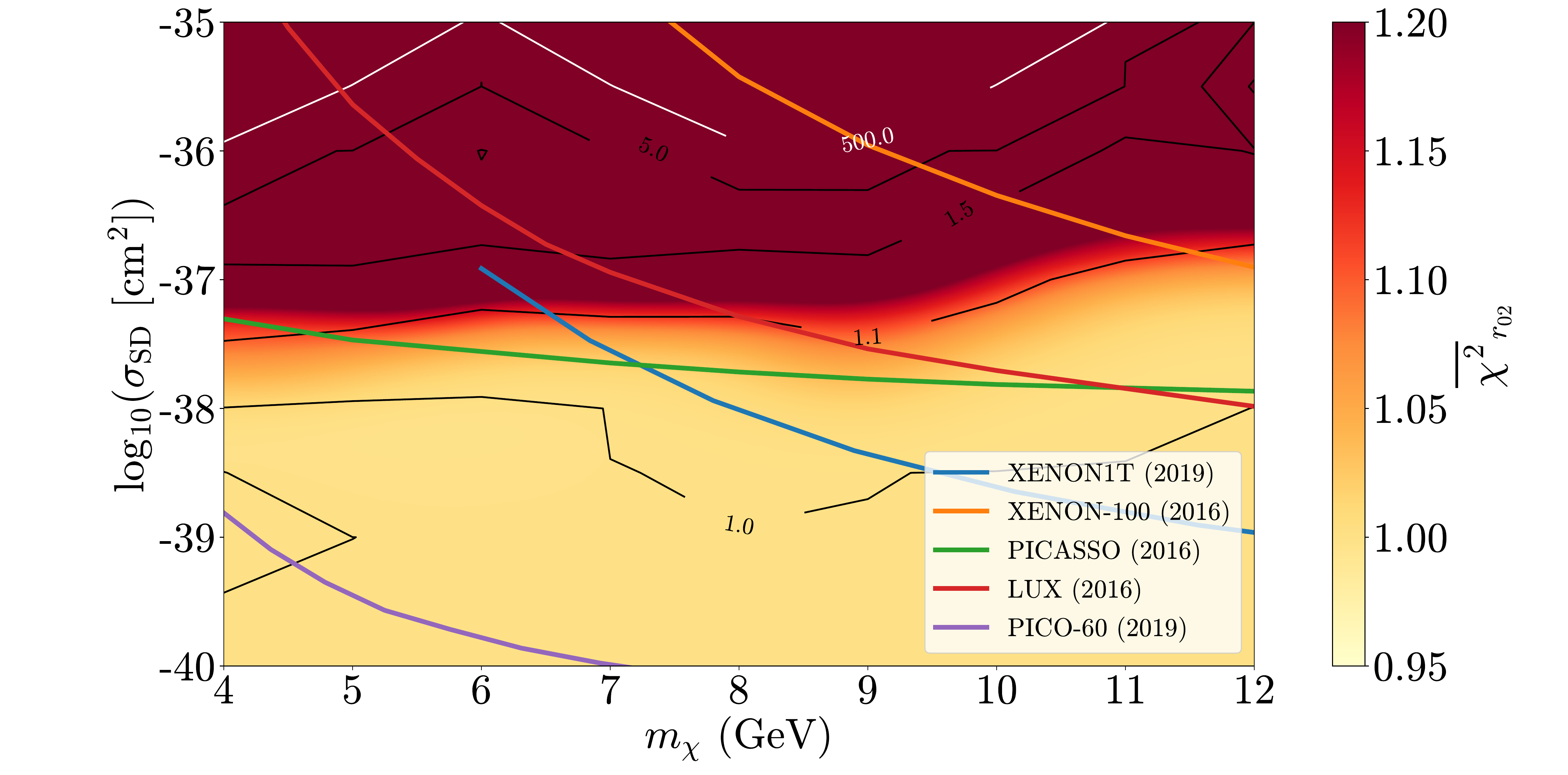}
    \caption{Contours for $\overline{\chi}^2_{r_{02}}$ (eq.~(\ref{eq:chi2r02}) and normalised similarly to eq.~(\ref{eq:chi2starnorm})): lighter shades represent lower $\chi^2$ and thus better models for this specific diagnostic. The contour for $\overline{\chi}^2_{r_{02}}=1$ represents models as good as the SSG model. Direct detection experiment upper bound limits from XENON1T (Aprile et al. \citeyear{XENON1T_2019}), XENON-100 (Aprile et al. \citeyear{XENON_2016}), LUX (Akerib et al. \citeyear{LUX}), PICASSO (Behnke et al. \citeyear{PICASSO_2017}) and PICO-60 (Amole et al. \citeyear{PICO_2019}) are shown as solid lines.}
    \label{fig:chi2r02}
\end{figure}

The results shown in Fig.~\ref{fig:chi2r02} confirm the overall trend seen before in Fig.~\ref{fig:chi2star2}: models in the lower region of the grid seem to more accurately agree with the observed $r_{02}$ which, in itself, grants more confidence to the previous results.
The contour that defines models with similar performance to that of the SSG model was again explicitly drawn at $\overline{\chi}^2_{r_{02}} = 1$.
Once more, most models with $\sigma_{SD} < 10^{-38}$ cm$^2$ outperform the best model with no DM interactions, this time on a different diagnostic that better represents the core structure.
It is also interesting to note that the best performing model of the previous set (Fig.~\ref{fig:chi2star2}) is not in the same region of the best performing models in the $\chi^2_{r_{02}}$ diagnostic.
This is a case where the spectroscopic results out-shadowed the structural differences in the first diagnostic, which was then covered by calculating the $r_{02}$ ratio.
Hence, the two step rejection method proves to be valuable in cases like this, since normally that model would have been accepted by passing the $\chi^2_\mathrm{star}$ diagnostic with the lowest value.

\begin{figure}
    \includegraphics[clip, trim=2.5cm 0cm 2.cm 0cm, width=\linewidth]{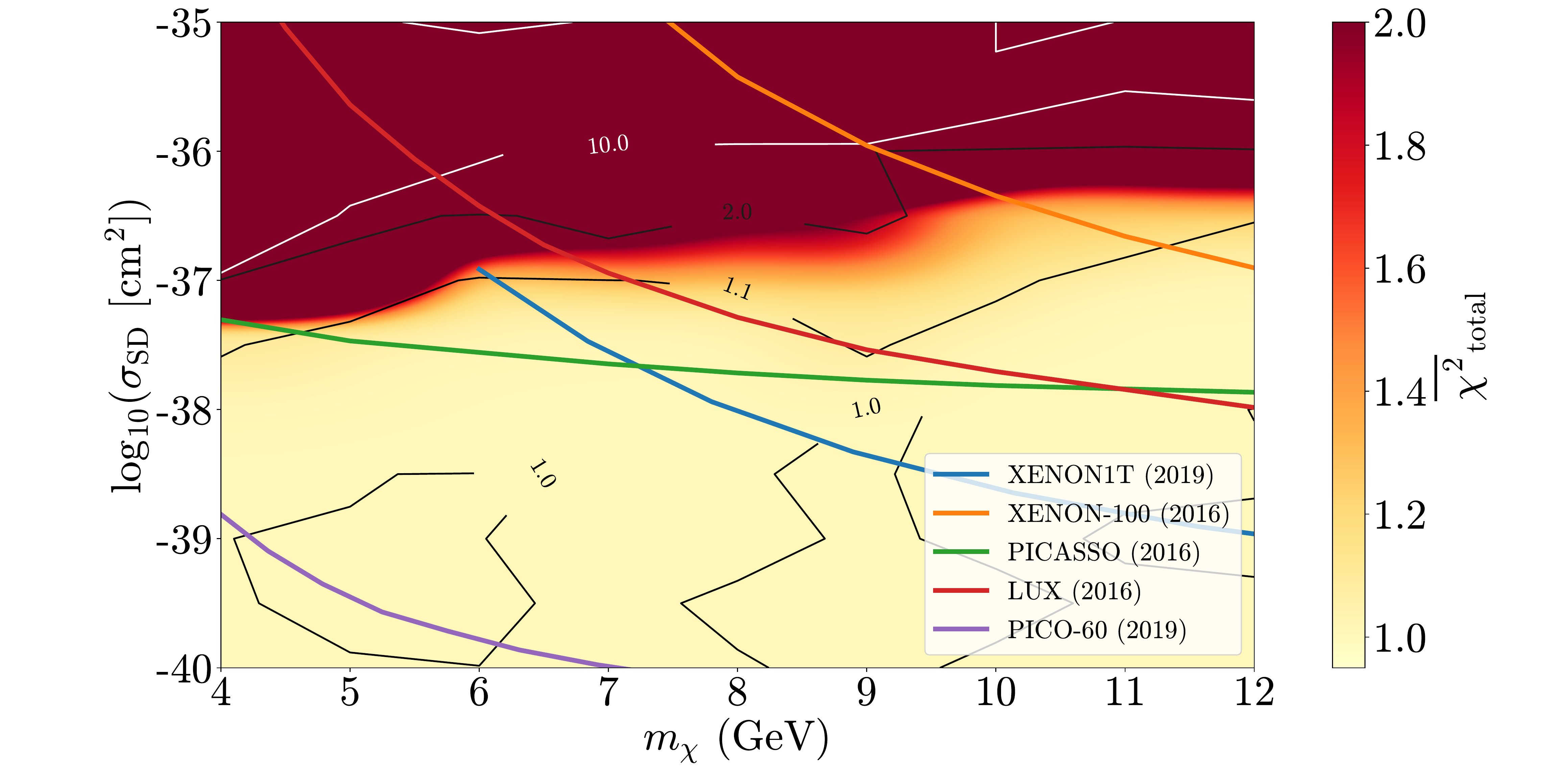}
    \caption{Maximum value between $\chi^2_\mathrm{star}$ and $\chi^2_{r_{02}}$: the contour for $\overline{\chi}^2_\mathrm{total} = 1$ represents the models with an equated performance to the SSG model, using their worst diagnostic. Direct detection experiment upper bound limits from XENON1T (Aprile et al. \citeyear{XENON1T_2019}), XENON-100 (Aprile et al. \citeyear{XENON_2016}), LUX (Akerib et al. \citeyear{LUX}), PICASSO (Behnke et al. \citeyear{PICASSO_2017}) and PICO-60 (Amole et al. \citeyear{PICO_2019}) are shown as solid lines.}
    \label{fig:chi2total}
\end{figure}

A simple additional test can be done by combining the two methods, taking the maximum value of both diagnostics, $\overline{\chi}^2_\mathrm{total} = \max(\overline{\chi}^2_\mathrm{star}, \overline{\chi}^2_{r_{02}})$.
This is shown in Fig.~\ref{fig:chi2total}, as well as the region within the $\overline{\chi}^2_\mathrm{total} = 1$ contour which represents the models that outperform the SSG model in both diagnostics.

When comparing the aforementioned grids with the direct detection experiment's limits we see that our method could provide complementary $m_\chi-\sigma_\mathrm{SD}$ exclusion diagrams.
This could be achieved by defining a cut-off $\chi^2$ since it is visible that there is a steep transition region between the two regimes (from yellow to dark red) which indicates a rapid disagreement between the stellar models and the observational data.
The behaviour showcased in Figs.~\ref{fig:chi2star2} - \ref{fig:chi2total} can be compared to that of the limits from PICASSO (\citeyear{PICASSO_2017}) and LUX (\citeyear{LUX}).
Moreover, it seems to suggest harder limits than those of XENON-100 (\citeyear{XENON_2016}), which is not as competitive for lower $m_\chi$.
On the other hand, the more recent XENON1T (\citeyear{XENON1T_2019}) and PICO-60 (\citeyear{PICO_2019}) report stronger limits for heavier masses, with the latter being more competitive.
Taking into account the SSG model as the benchmark we can hint towards the region with $\sigma_\mathrm{SD} \gtrsim 10^{-37}~\mathrm{cm}^2$ where $\overline{\chi}^2_\mathrm{total}$ rapidly increases, meaning that ADM presence in this SG star is strongly disfavoured by both spectroscopic and seismic observations.

Finally, it should be noted that the models obtained in the previous section -- which were obtained by calibrating the stellar and, in the case of DM Calib, the DM parameters -- can fall in the excluded region suggested in Fig.~\ref{fig:chi2total} (see Table \ref{table:benchmark}) and still yield stellar models that are in agreement with the observational data.
This may be due to the increase in the degrees of freedom associated with the extra free parameters which, given that the observational error for the star in question is substantial, allows the method to find different combinations of parameters that still fit the reality.
Therefore, with access to more precise measurements from future spectroscopy and asteroseismology missions, one can expect to calibrate a standard benchmark model that allows the drawing of exclusion diagrams with more certainty.

\subsection{Period Spacing Analysis}

When further analysing the oscillation eigenfunctions of several models, it is clear that the amplitude rapidly falls off within the first outer 20\% of the star radius.
This means that despite using a seismic ratio that is designed to gather more information about the stellar interior, the acoustic modes that defined the diagnostic fall short on this task.
Thus, we can conclude that the $r_{02}$ diagnostic is not as sensitive to the core as expected, being more representative of the stellar envelope, contrary to what happens in a typical MS star.

Motivated by this, we change our focus to the gravitational character of the oscillations, which, as seen before, should be specially sensitive to the stellar interior.
This is done by analysing the mixed modes of the various models, which allows the extraction of the relevant quantities of the gravity contribution.
In particular, the asymptotic value of the large period separation presented in eq.~(\ref{eq:DeltaPi}) can be computed from the MESA models for $\ell=1$ and $\ell=2$, for which there are mixed modes.
It is expected that if the presence of DM in stars directly affects the size of their convective cores, as we have seen in previous sections, then the period spacing will serve as a good probing tool (Lopes et al. \citeyear{Lopes_2019b}).
The $\Delta\Pi_\ell$ values are shown in the two last columns of Table \ref{table:benchmark}.
The already mentioned regimes are once again clearly noticeable: there is a substantial decrease in period spacing when DM is strongly influential that can go up to 66\% of the benchmark value (from SSG).
This, again, is explained by the suppression of the convective core caused by DM, which in turn leads to an increase of the g-mode cavity, producing a direct effect on the integral in eq.~(\ref{eq:DeltaPi}).

\begin{figure}
    \includegraphics[clip, trim=2.5cm 0cm 2.5cm 0cm, width=\linewidth]{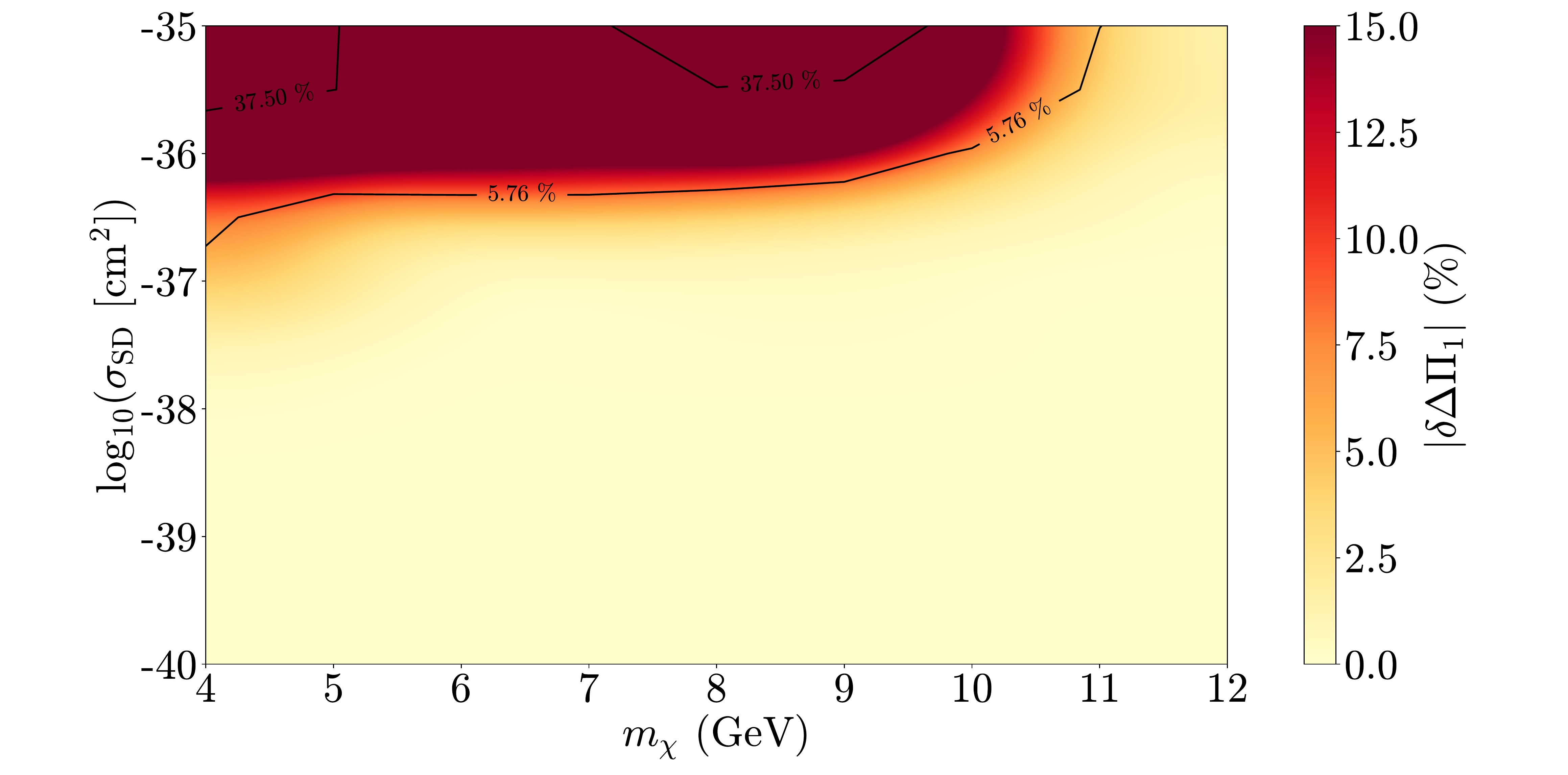}
    \caption{Contour plot of the deviation of the period separation $\Delta\Pi_1$ from the SSG model for the same grid of DM models. The two black curves represent the lower (5.76 per cent) and upper (37.50 per cent) boundaries of the relative error in the measurement of $\Delta\Pi_1$ (proportion of the uncertainty in relation to the measured value) in a set of 39 SG stars from Mosser et al. (\citeyear{Mosser_2014}) (Lopes et al. \citeyear{Lopes_2019b}).}
    \label{fig:deltaDPi_l1}
\end{figure}

As before, we define the $\Delta\Pi_\ell$ deviation from the SSG model for the same grid of DM models to better understand the impact that the different DM parameters have on the stellar core.
The variation of $\Delta\Pi_1$ is shown in Fig.~\ref{fig:deltaDPi_l1}.
Only the case with $\ell=1$ is shown, given that, by definition, in the asymptotic limit, $\Delta\Pi_1$ and $\Delta\Pi_2$ only differ by a multiplying constant.
The darker region represents models where $|\delta\Delta\Pi_1|$ is larger than 15\% and, in some cases, models reach the 60\% mark as was already the case for some models in Table \ref{table:benchmark}.
It is also clear that larger masses of ADM particles tend to result in less effects.
This is twofold: first, these particles are harder to capture by virtue of requiring a larger transfer of momentum upon recoil in order to reach a velocity lower than the escape velocity (e.g. Gould \citeyear{Gould1987}); secondly, if eventually captured, they cluster strongly in the innermost regions of the star and, thus, their impact is naturally not felt as much.
It is important to note that most of the deviations shown in the grid happen in the negative direction, with DM impacted models exhibiting a smaller period spacing, as expected.

The overall tendency in Fig.~\ref{fig:deltaDPi_l1} mimics that of Figs.~\ref{fig:chi2star2} - \ref{fig:chi2total}: models in the upper part of the grid show a larger ADM influence on the star.
Yet again, the transition region between the two regimes is narrow.
The region below the 5.76 per cent contour represents models whose $\Delta\Pi_1$ variation is lower than the lowest relative uncertainty among the large period separation measurements in Mosser et al. (\citeyear{Mosser_2014}).
Likewise, models above the 37.50 per cent curve exhibit a variation that surpasses the uncertainty of the less precise measurements.
Both these statements mean that current experiments may not have enough sensitivity to resolve DM signatures in SG stars in the transition region (where the lower relative uncertainty contour is) and particularly in the region below the 37.50\% contour.
Above that curve, the effect should be detected, with the deviation being greater than the observable uncertainty in the worst cases. However, one should note that if this analysis is carried out for a SG star which allows for a more precise measurement of $\Delta\Pi_1$ (i.e., closer to the 5.76\% mark) there should be enough sensitivity to draw any valuable conclusions regarding the acceptability of DM models for that star.
In the future, this diagnostic could be used with measurement data as the benchmark, providing a strong case for model rejection and, additionally, ADM parameter constraints.

\section{Conclusions}

Subgiant stars are but a small fraction of the currently observed stars by virtue of that evolutionary stage being relatively short.
Despite that, they can serve as important laboratories for the study of DM constraints.
In this work, we obtained calibrated stellar models of the subgiant KIC 8228742 assuming both the presence and absence of asymmetric DM spin-dependent interactions.
Focusing on SD interactions allows us to directly compare our results to the corresponding constraints placed by detectors which study the same type of interactions.
More importantly, by studying only spin dependent couplings, we essentially neglect interactions with elements heavier than hydrogen, and thus also any direct effect that DM has on the star during the SG phase. %rendering the SG phase as non-contributory to the ADM signatures in the star.
%This means that, in fact, we neglect any DM energy transport that occurs during the actual SG phase, given that in these stars the region inhabited by DM particles, i.e., the core, is mostly devoid of hydrogen.
%Thus, the DM signatures studied in this work -- and the resulting departures from the standard non-DM models -- are in fact a consequence of DM interactions that occurred mainly during the MS.
%By carrying out this study we take advantage of the seismological benefits of the SG phase to study DM interactions in the MS stage.
Then, by carrying out this study we take advantage of the seismological benefits of the SG phase to study DM interactions in the MS stage.

The results shown here point towards the fact that, overall, DM models are in agreement with current observations of this star.

Firstly, in an attempt to study the possibility of ADM presence in SG stars, we introduce DM parameters into the calibration and optimisation processes.
This is a new approach which aims to find the best models, in terms of the diagnostics here proposed, disregarding any prior standard (with no DM influence) benchmark models.
Calibrated models with strong DM influence showcased a different regime from both the standard (SSG) model and DM models with lower $\sigma_\mathrm{SD}$.
Phenomena like the suppression of the convective core, the cooling of the inner core and the increase in density of that region were all present and in agreement with past findings (e.g. Taoso et al. \citeyear{Taoso}; Iocco et al. \citeyear{Iocco_2012}; Casanellas \& Lopes \citeyear{Casanellas2013}; Casanellas et al. \citeyear{Casanellas2015}; Lopes \& Silk \citeyear{Lopes_2002}).
More recently, Raen et al. (\citeyear{Raen_2021}) studied the effects of ADM on the stellar evolution of 0.9 - 5 M$_\odot$ stars for $\sigma_\mathrm{SD} = 10^{-37}$ cm$^2$ and higher DM densities ($\rho_\mathrm{DM} \sim 10^2 - 10^5$ GeV/cm$^3$).
Their findings align with the results of this work, having reported a generalised suppression of convection in the cores of stars with masses $M \gtrsim$ 1.3 $M_\odot$ added to a flattening of the temperature profile in the core and an increase in MS lifetimes by up to 20\% for stars with $0.9 \leqslant M/M_\odot\lesssim 1.3$.
We later conclude that the results obtained at this stage of the work, which provided an optimal pair of ADM particle mass and spin-dependent cross-section of $m_{\chi} = 9.12~\mathrm{GeV}$ and $\sigma_{\mathrm{SD}} = 2.32 \times 10^{-36} \mathrm{cm}^2$ may not be sufficient to constrain the DM particle candidates' properties.
A number of limitations can be pointed out as the cause for this. The number of spectroscopic quantities of this star that we have measurements for pose a problem when introducing more parameters into the input group, since overfitting -- typical of situations where there are too many parameters relative to the available data points -- may be happening here.
The precision of the current measurements is also a factor since it broadens the accepted model spectrum and limits the certainty of a possible exclusion limit.
Lastly, computational time constraints were a large limitation to the minimisation problem and this is an aspect that has the potential for clear improvements.

Using seismological diagnostics as a second probing tool, we then present a method to study the influence of DM in the interior of stars, with direct applicability to SGs.
The $r_{02}$ ratio is used in an attempt to probe the stellar core, which is the region that ADM severely impacts.
A study of the deviation of $r_{02}$ from observational measurements allows us to draw several sensitivity grids (Figs.~\ref{fig:chi2star2} - \ref{fig:chi2total}) which showcase the increasing influence that DM has with the increase of the spin-dependent cross-section of the interaction between ADM particles and stellar matter.

Moreover, those same figures show a class of models with $10^{-40}\leqslant\sigma_\mathrm{SD}<10^{-38}$ cm$^2$ that consistently outperform the best standard models, which strengthens the argument that the presence of ADM particles in this star is consistent with observations.
Additionally, a $\sigma_\mathrm{SD}$ admissible region is suggested for values up to 10$^{-37}$ cm$^2$ where the $\chi^2$ diagnostics start to deteriorate from observations.
This value is comparable to those of the PICASSO (\citeyear{PICASSO_2017}) and LUX (\citeyear{LUX}) experiments and improves that of XENON-100 (\citeyear{XENON_2016}), whilst not being able to compete with the bounds set by PICO-60 (\citeyear{PICO_2019}).
Then, an exclusion limit may be drawn with more certainty at $\sigma_\mathrm{SD} \gtrsim 10^{-35}$ cm$^2$ since results from both Section~\ref{sec:stellarmodels} and \ref{sec:discussion} find this region as incompatible with current observations.
This conclusion is similar to what Casanellas \& Lopes (\citeyear{Casanellas2013}) and Martins et al. (\citeyear{Martins_2017}) found previously for MS stars.

Lastly, we propose an additional seismic parameter to study DM influence on SG stars that allows us to further probe the stellar core.
As the acoustic modes were mainly probing the stellar envelope (contrary to what typically happens in MS stars), we turned our attention to gravity and mixed modes, which carry more information about the core since they travel deeper into the stellar interior.
The $\Delta\Pi_\ell$ diagnostic was calculated for the same grid of models and the results again confirm the previous analysis and hint towards the possibility of drawing exclusion diagrams for a SG star for which we have the $\Delta\Pi_\ell$ observations.

In the future, missions like PLATO (Rauer et al. \citeyear{PLATO}), which will provide high precision measurements of both spectroscopic and seismic quantities, will allow for a more extensive analysis of the impact of ADM in stars and, more importantly, will make the drawing of exclusion diagrams possible, using the method we present here.
With better measurements, we also expect the calibration with DM quantities as inputs to provide a better result, with less impact from numerical constraints.

Finally, we should note that we only considered DM with spin-dependent interactions in this work, which only account for scatterings with hydrogen. Therefore, one aspect to consider in the future is the inclusion of spin-independent interactions, which, although more limited by direct detection, should have a different impact in the later stages of stellar evolution, when stars form cores made of helium and heavier elements.

\section*{Acknowledgements}

We thank Bill Paxton and all MESA contributors along with J\o rgen Christensen-Dalsgaard (ADIPLS: Christensen-Dalsgaard \citeyear{ChristensenDalsgaard2008}) and Rich Townsend (GYRE: Townsend \& Teitler \citeyear{Townsend2013}; Townsend, Goldstein \& Zweibel \citeyear{Townsend2018}) for making their code and work publicly available. Additionally, we acknowledge the \href{https://particleastro.brown.edu/}{Brown University's Particle Astrophysics Group} for maintaining the \href{http://dmtools.brown.edu/limits}{DMTools} database and plotting resources.
We also thank the anonymous referees for the valuable comments and suggestions.

J.L. acknowledges financial support from Fundação para a Ciência e Tecnologia (FCT) grant No. PD/BD/128235/2016 in the framework of the Doctoral Programme IDPASC---Portugal.

I.L. thanks the Fundação para a Ciência e Tecnologia (FCT), Portugal, for the financial support to the Center for Astrophysics and Gravitation (CENTRA/IST/ULisboa) through the Grant Project No. UIDB/00099/2020 and Grant No. PTDC/FIS-AST/28920/2017.
%%%%%%%%%%%%%%%%%%%%%%%%%%%%%%%%%%%%%%%%%%%%%%%%%%
\section*{Data Availability}

This article makes use of the MESA stellar evolutionary code that is available for free in the public domain and can be found at the Zenodo repository with the DOI: 10.5281/zenodo.3473377.
All of the data sets used in this work can be found in the MESA files for the release version r12115.
The GYRE code has also been used to compute oscillations, and is also available for free in \url{https://github.com/rhdtownsend/gyre} with the corresponding documentation in \url{https://gyre.readthedocs.io/}.
The data sets for the exclusion limits of the direct detection experiments shown in this work are publicly available upon registration in \url{http://dmtools.brown.edu/}.
The remaining data underlying this article will be shared on reasonable request to the corresponding author.



\bibliographystyle{mnras}
\bibliography{references} % if your bibtex file is called example.bib

%\begin{thebibliography}{99}
%\end{thebibliography}

%%%%%%%%%%%%%%%%%%%%%%%%%%%%%%%%%%%%%%%%%%%%%%%%%%

%%%%%%%%%%%%%%%%% APPENDICES %%%%%%%%%%%%%%%%%%%%%

%\appendix



%%%%%%%%%%%%%%%%%%%%%%%%%%%%%%%%%%%%%%%%%%%%%%%%%%

\vspace{6mm}
% Don't change these lines
\bsp	% typesetting comment
\label{lastpage}
\end{document}